\documentclass[12pt]{article}

\usepackage{sbc-template}
\usepackage{graphicx,url}
\usepackage[utf8]{inputenc}
\usepackage[misc]{ifsym}
     
\title{PhysiBoSS-Models: A database for multiscale models}

\author{%
Vincent Noël\inst{1,2,3}\Letter, %
Marco Ruscone\inst{4}, %
Randy Heiland\inst{6}, %
\\%
Arnau Montagud\inst{4,5}, %
Alfonso Valencia\inst{4, 6}, %
\\%
Emmanuel Barillot\inst{1,2,3}, %
Paul Macklin\inst{6}, %
Laurence Calzone\inst{1,2,3}\Letter%
}

\address{
Institut Curie, PSL Research University, Paris, France
\nextinstitute
INSERM, U1331, Paris, France
\nextinstitute
MINES ParisTech, CBIO-Centre for Computational Biology, Paris, France
\nextinstitute
Barcelona Supercomputing Center, Barcelona, Spain
\nextinstitute
Institute for Integrative Systems Biology (I2SysBio), CSIC-UV, Valencia, Spain
\nextinstitute
ICREA, 23 Passeig Lluís Companys, 08010 Barcelona, Spain
\nextinstitute
Department of Intelligent Systems Engineering,\\Indiana University. Bloomington, IN, USA
\email{vincent.noel@curie.fr, laurence.calzone@curie.fr}}

\begin{document} 

\maketitle

\begin{abstract}
  PhysiBoSS is an open-source platform that integrates agent-based modeling of cell populations with intracellular stochastic Boolean networks, enabling multiscale simulations of complex biological behaviors. To promote model sharing and versioning, we present the PhysiBoSS-Models database: a curated repository for multiscale models built with PhysiBoSS. By providing a simple Python API, PhysiBoSS-Models provides an easy way to download and simulate preexisting models through tools such as PhysiCell Studio. By providing standardized access to validated models, PhysiBoSS-Models facilitates reuse, validation, and benchmarking, supporting research in biology.
\end{abstract}
     


\section*{Introduction}

The modeling of biological systems increasingly requires the integration of processes occurring on multiple spatial and temporal scales\cite{metzcar_review, Cappuccio, MONTAGUD2021100385}. From intracellular molecular signaling to cell population dynamics and tissue-level interactions, multiscale models provide a powerful framework to capture the complexity of biological behaviors. Among the tools developed for this purpose, PhysiBoSS\cite{letort2019physiboss,ponce2023physiboss} has emerged as a versatile open-source platform that couples agent-based modeling with stochastic continuous-time Boolean networks to simulate the decision-making processes of individual cells within their microenvironment.

Despite the growing use of PhysiBoSS in fields such as cancer biology and immunology, the community still faces important challenges regarding model sharing, model versioning and collaborative development. Models can and should be built and modified from existing models to take advantage of the work already done by the community. Currently, PhysiBoSS models are often stored in project-specific repositories, with missing instructions for compilation or information about compatibility with other tools, limiting their reuse and validation by other research groups.

To address this gap, we introduce the PhysiBoSS-Models database, a curated and standardized resource designed to store, organize, and disseminate multiscale models developed with the PhysiBoSS framework. PhysiBoSS-Models uses GitHub to store models, providing powerful version control and enabling automatic testing-deployment workflows, and is available as \url{https://github.com/PhysiBoSS-Models}. PhysiBoSS-Models aims to promote reusability, facilitate model exchange, and accelerate the adoption of multiscale modeling approaches across the systems biology community.

\section{PhysiBoSS}
PhysiBoSS was developed to address the need for simulating biological systems in which cell-level decision-making depends on complex intracellular signaling, while also accounting for interactions between individual cells and their surrounding environment. The integration of these two scales is achieved by coupling two modeling approaches:

\subsection{Agent-Based Modeling}
    This layer captures the behavior and interactions of individual cells in a 2D/3D spatial environment. It allows simulation of processes such as cell proliferation, migration, death, and response to external stimuli. PhysiBoSS is an add-on of PhysiCell\cite{ghaffarizadeh2018physicell}, a widely used agent-based modeling engine for multicellular systems. To simplify its usage, PhysiCell comes with a graphical user interface, PhysiCell Studio\cite{heiland2024physicell}, which is fully compatible with PhysiBoSS.

\subsection{Stochastic Boolean Modeling}
    At the intracellular level, PhysiBoSS incorporates MaBoSS\cite{stoll2017maboss}, a continuous-time stochastic Boolean framework, to model the dynamics of gene regulatory and signaling pathways within each cell. This allows the representation of complex signaling behaviors without the computational burden of fully quantitative models, making it suitable for large cell populations. 


However, such multiscale models are very complex to use and to modify as this involves making custom C++ code. Providing an easy and standardized way to share these models, including cross-platform precompiled binaries, would greatly facilitate their reuse.

\section{Database Design and Implementation}

PhysiBoSS-Models has been designed to provide a version-controlled repository for multiscale models built with the PhysiBoSS framework. To facilitate this aspect, PhysiBoSS-Models is based on Git and hosted on GitHub, leveraging its robust version control system for tracking model updates, managing community submissions, and exploiting its continuous integration infrastructure.

\subsection{Repository Structure and Content}

Each model entry in PhysiBoSS-Models follows a standardized folder structure of a PhysiBoSS model, including the core PhysiBoSS and PhysiCell source code which allows modelers to pin their model to a specific version of the framework. The model can also include custom C++ modules that provide specific functionalities. Besides code, the model contains configuration files such as the XML settings, cell positions, cell rules\cite{johnson_cell_grammar_2025_short}, or intracellular Boolean models. Finally, modelers provide a model description in the readme file of the repository and a metadata file (model.yml) giving structured information such as model version and the list of different configuration files. 


Upon submission of a new model, or for a new version of an existing model, these entries are compiled and packages are generated for Windows, macOS and Linux operating systems. 

\subsection{Version Control and Community Contributions}

By hosting the database on GitHub, PhysiBoSS-Models benefits from built-in version control, allowing precise tracking of model updates, bug fixes, and improvements. Researchers can submit pull requests to contribute to new models or propose updates, while issue tracking enables discussion and resolution of potential problems or inconsistencies.

Each model release is tagged to ensure proper citation and reproduce specific historical versions of a model used in previous publications.



\begin{figure}[ht]
\centering
\includegraphics[width=.8\textwidth]{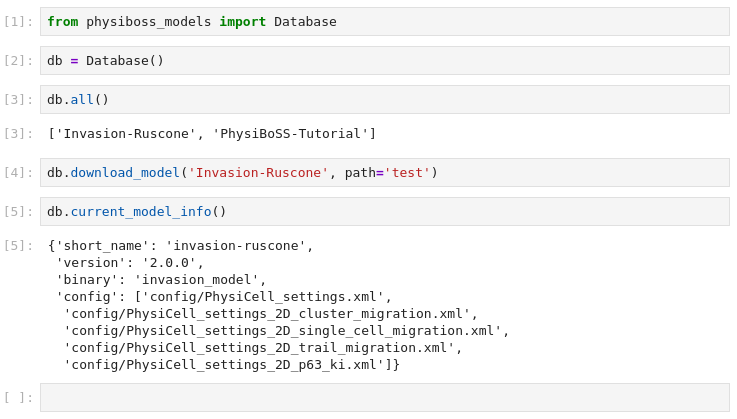}
\caption{Example of usage of the Python API, within a Jupyter notebook}
\label{fig:python_api}
\end{figure}

\subsection{Programmatic Access via Python API}

In addition to manual download, PhysiBoSS-Models provides a Python-based API that allows users to search and retrieve models programmatically, based on keywords. Models can then be downloaded and unpacked directly into user-defined directories. Finally, model metadata is accessible for automated workflows in larger modeling pipelines. 

An example of usage is shown in Figure \ref{fig:python_api}. This API promotes seamless integration of PhysiBoSS models into computational pipelines. 

\subsection{PhysiCell Studio Integration}

\begin{figure}[ht]
\centering
\includegraphics[width=.9\textwidth]{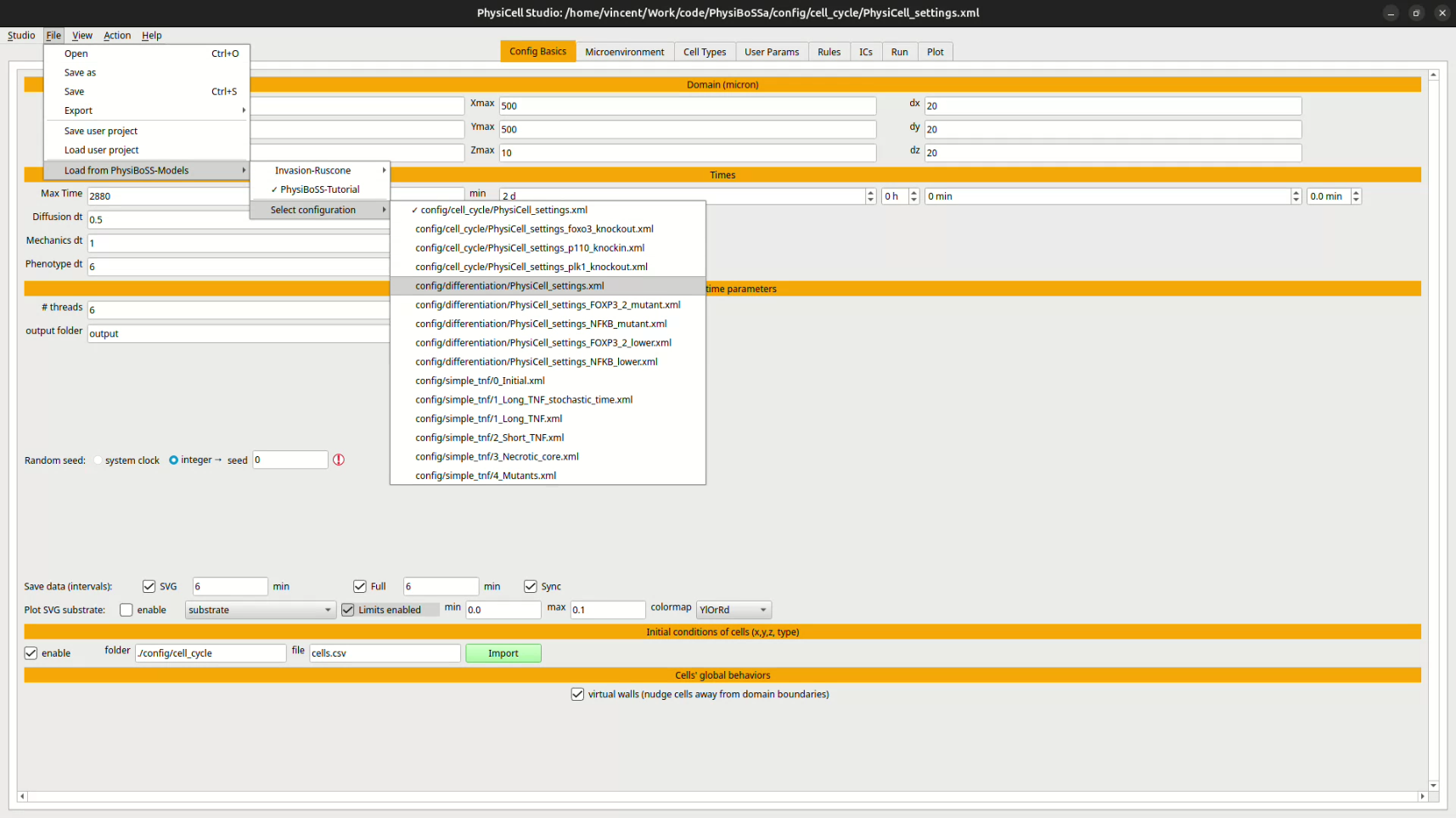}
\caption{Screenshot of PhysiCell Studio, showing the selection of a specific configuration file of the PhysiBoSS tutorial.}
\label{fig:physicell_studio_integration}
\end{figure}

To make it easily available to most users, we implemented support for PhysiBoSS-Models in PhysiCell Studio as seen in Figure \ref{fig:physicell_studio_integration}.

This allows most users to use PhysiBoSS-Models and run published models in a few clicks, without technical requirements, which historically limited the access to such models. 

\section{Example Models}

We started PhysiBoSS-Models with two peer-reviewed models: the demonstration models from the PhysiBoSS tutorial\cite{ruscone2024building} and a cancer invasion model\cite{ruscone2023multiscale}.

\subsection{PhysiBoSS tutorial models}

The PhysiBoSS tutorial, published in 2024, showcases the new automated mapping functionalities between PhysiBoSS and PhysiCell. It comes with three simple examples demonstrating different aspects of PhysiBoSS: a TNF-mediated tumor death model shows how different treatments can lead to different fates, a mechanistic cell cycle model controlled by a Boolean model, and a T-cell differentiation model illustrating the cell-cell interaction leading to T cell differentiation. 
Since these three models can run with the same PhysiBoSS binary, we packaged them together, presenting an example of multiple configuration files accessible in the same model.

\subsection{Cancer invasion model}
This model studies the influence of both physical parameters and intracellular signaling pathways on the different modes of cancer invasion (f.i., single or collective migration) and was first published in 2023. A new version of this model was then published in the PhysiBoSS tutorial in 2024. Both versions are accessible in the database under versions 1.0.0 and 2.0.0, respectively. This model is a good example of how versioning can be handled with PhysiBoSS-Models. 

This list is growing as new models are published. The procedure to submit models is described on the website. Editors will verify that the models are compatible with PhysiBoSS, and will be added to the list.

\section{Discussion and Future Perspectives}

The development of PhysiBoSS-Models represents an important step toward improving the sharing of already-published models, transparency, and collaboration within the multiscale modeling community. By providing a centralized, version-controlled repository with standardized formats and comprehensive metadata, PhysiBoSS-Models addresses many of the barriers that have historically limited the sharing and reuse of complex models in systems biology.

One of the key strengths of PhysiBoSS-Models lies in its integration with GitHub, which not only ensures long-term sustainability through community-driven maintenance but also provides full version tracking for each model. This allows researchers to refer to specific model versions used in published studies, enhancing scientific rigor, and facilitating independent validation.

The inclusion of cross-platform binaries, removing the need for complex compilation steps, further lowers the entry barrier for new users. 
Notably, the integration with PhysiCell Studio allows them to implement new models in minutes, instead of hours.
Moreover, the Python API extends the usability of PhysiBoSS-Models by allowing researchers to integrate PhysiBoSS models into larger computational workflows, including automated parameter sweeps, sensitivity analyses, and visualization pipelines.

Looking ahead, several future developments are planned to expand the scope and impact of PhysiBoSS-Models. First, we want to extend our database beyond PhysiBoSS to include any model built with PhysiCell. We also want to produce simulation outputs via our automatic workflow, allowing users to easily access reference results. 
Finally, we plan to provide a quality web interface for browsing models, exploring simulation outcomes and metadata, and launching simulations of future web-based platforms. 

    
Ultimately, PhysiBoSS-Models platform aims to become a reference resource for multiscale modeling in systems biology, fostering a more open and collaborative research environment. By making high-quality models easily accessible and reusable, we hope to accelerate the development and adoption of predictive simulations in areas such as cancer research, immunology, and tissue engineering.



\section*{Funding}
This work has been supported by the French Plan Cancer-MIC ITMO, Project MoDICeD (No.20CM114-00) and by the French Agence Nationale pour la Recherche (ANR) under the France 2030 program, with the reference number ANR-24-EXCI-0004. AM acknowledges funding from the Generalitat Valenciana’s CIDE GenT programme under the project CIDEXG/2023/22. MR acknowledges funding from the Ayuda Severo Ochoa CEX2021-001148-S from MICIU/AEI.



\bibliographystyle{sbc}
\bibliography{sbc-template}

\end{document}